\documentclass[a4paper,twoside]{article}

\usepackage{epsfig}
\usepackage{subcaption}
\usepackage{calc}
\usepackage{amssymb}
\usepackage{amstext}
\usepackage{amsmath}
\usepackage{amsthm}
\usepackage{multicol}
\usepackage{pslatex}
\usepackage{graphicx}
\usepackage[backgroundcolor=blue!10,bordercolor=gray]{todonotes}

\usepackage{footnote}
\usepackage{url}
\usepackage{apalike}
\usepackage{SCITEPRESS}     

\graphicspath{{resources/}}

\begin{document}

\title{Improving Vulnerability Prediction of JavaScript Functions Using Process Metrics}

\author{\authorname{Tamás Viszkok\sup{1}\orcidAuthor{0000-0002-6049-786X}, Péter Hegedűs\sup{1,2}\orcidAuthor{0000-0003-4592-6504} and Rudolf Ferenc\sup{1,3}\orcidAuthor{0000-0001-8897-7403}}
\affiliation{\sup{1}Department of Software Engineering, University of Szeged, Dugonics tér 13, Szeged, Hungary}
\affiliation{\sup{2}MTA-SZTE Research Group on Artificial Intelligence, ELKH, Tisza Lajos krt. 103, Szeged, Hungary}
\affiliation{\sup{3}FrontEndART Ltd., Somogyi utca 19., Szeged, Hungary}
\email{\{tviszkok, hpeter, ferenc\}@inf.u-szeged.hu}
\vspace{-35pt}}

\keywords{Vulnerability prediction, static source code metrics, process metrics, JavaScript security}

\abstract{Due to the growing number of cyber attacks against computer systems, we need to pay special attention to the security of our software systems.
In order to maximize the effectiveness, excluding the human component from this process would be a huge breakthrough.
The first step towards this is to automatically recognize the vulnerable parts in our code.
Researchers put a lot of effort into creating machine learning models that could determine if a given piece of code, or to be more precise, a selected function, contains any vulnerabilities or not.
We aim at improving the existing models, building on previous results in predicting vulnerabilities at the level of functions in JavaScript code using the well-known static source code metrics.
In this work, we propose to include several so-called process metrics (e.g., code churn, number of developers modifying a file, or the age of the changed source code) into the set of features, and examine how they affect the performance of the function-level JavaScript vulnerability prediction models. 
We can confirm that process metrics significantly improve the prediction power of such models.
On average, we observed a 8.4\% improvement in terms of F-measure (from 0.764 to 0.848), 3.5\% improvement in terms of precision (from 0.953 to 0.988) and a 6.3\% improvement in terms of recall (from 0.697 to 0.760).
\vspace{-10pt}}

\onecolumn \maketitle \normalsize \setcounter{footnote}{0} \vfill

\section{\uppercase{Introduction}}
\label{sec:introduction}
\vspace{-5pt}

Nowadays, thanks to the advances in technology, software is no longer only available in the traditional computing environments.
It is in our pockets on smart-phones, in the kitchen in our refrigerators, and it is even there in the smart-watches worn on our wrists.
As these devices become an integral part of our daily lives, our vulnerability to cyber-crimes is growing fast.
Therefore, the need to create secure software and detect software vulnerabilities as effectively as possible is growing rapidly as well.

Nevertheless, in many cases, companies do not devote sufficient resources to detect such vulnerabilities due to tight deadlines, or may even completely neglect and entrust this task to developers.
But even for companies that employ a dedicated security professional to make sure that their software is protected against cyber-crimes, the human factor can still cause a problem.
The introduction of an automated method would help to solve this issue.
Moreover, the sooner a software vulnerability is detected, the greater the amount of the saved resources that would otherwise be wasted.
Perhaps more importantly, the release of a software that contains vulnerabilities can significantly damage the reputation of developers, but even more so that of the company.
In addition, such cases could even lead to lengthy lawsuits, which we ideally want to avoid.
Therefore, it would be invaluable if a tool could detect whether a particular piece of code contains any possible vulnerabilities before they are released into production.

Fortunately, we are living in the heyday of artificial intelligence.
The high-computational capacity needed to train learning algorithms is becoming more and more affordable, and the vast amount of data available to anyone on the Internet provides the perfect opportunity for algorithms requiring larger training datasets.
We can also take advantage of the capabilities of machine learning in the field of cyber-security.
Most of the existing error detection models focus mainly on predicting software defects in general.
However, in many cases, vulnerabilities cannot be considered as a defect in the traditional sense, so these methods cannot be applied with sufficient effectiveness against vulnerabilities without adjustment~\cite{7180086,sva,5477059,Shin2011CanTF}.

Our goal is to create an AI driven, precise, prediction model that works at a fine-grained level, namely to detect whether a given function of a JavaScript software contains any vulnerabilities or not.
The approach we take is to train a new, specialized model on a carefully assembled dataset, which contains vulnerable program samples (i.e., code that has been later fixed for a vulnerability) for training models.
As a starting point for our studies, we selected an already existing vulnerability dataset published by Ferenc et al.~\cite{raise} that contains 12,125 JavaScript functions with class labels if they are vulnerable or not.
This dataset contains 42 static source code metrics as features.
Although their initial results were already promising, due to the very dynamic properties of JavaScript we hypothesized that improvements could be made by extending the set of features for the prediction.

Given that general prediction models leveraged a lot from the addition of so-called process metrics~\cite{proc_met,proc_better,8073757,8004363} as features, we decided to augment the existing dataset~\cite{raise} with process metrics, and evaluate how this affects the performance of vulnerability prediction models.
Process metrics, such as \textit{Number of Modifications}, \textit{Age} or \textit{Number of Contributors} provide an additional aspect of the source code compared to its structural static code metrics.
For example, if a program element was modified many times in the past that might suggest some inherent problems with the code.
Whether being hard to change, error-prone, or simply too coupled with other components, all these properties make it suspicious from a security point of view.
After a refactoring of such code, it may be less likely to remain vulnerable.
The refactoring also resets the aforementioned process metrics, so our hypothesis is that they would properly capture properties of code that help predict their vulnerability.


To summarize our research goals, we formalized the following two research questions:

\textbf{RQ1}: Can process metrics as features improve existing JavaScript vulnerability prediction models based only on static source code metrics?

\textbf{RQ2}: If process metrics do improve the performance of vulnerability prediction models, how significant it is in terms of precision, recall, and F-measure?

By adding 19 process metrics to the set of features and re-training the prediction models we found that a significant increase could be achieved in model performances.
On average, we observed a 8.4\% improvement in terms of F-measure (from 0.764 to 0.848), 3.5\% improvement in terms of precision (from 0.953 to 0.988) and a 6.3\% improvement in terms of recall (from 0.697 to 0.760).

The rest of the paper is organized as follows.
In Section~\ref{sec:related}, we present the works related to our topic of research.
We give an overview of our applied methodology and process metric calculation in Section~\ref{sec:approach}.
The results of the new, extended vulnerability prediction models are detailed in Section~\ref{sec:results}, where we also compare these with the static source code based predictions.
Section~\ref{sec:threats} enumerates the list of possible threats, while we conclude the paper in Section~\ref{sec:conclusion}.

\vspace{-10pt}
\section{\uppercase{Related Work}}
\label{sec:related}
\vspace{-5pt}

The viability of error detection using artificial intelligence has already been demonstrated in many previous works~\cite{8816965,sbpmla,anmlabp,aessebbpuml,Tth2016APB,iceis19}.
However, these solutions are used to predict any kinds of errors that may prevent the program from functioning properly, and not only vulnerabilities.
But, in many cases, vulnerabilities do not fall into this generalized ``bug'' category.
This is why we decided to train a new model specifically designed for detecting vulnerabilities instead of using existing fault detection models.
Our approach of dedicated vulnerability models is shared by many researchers, as we can find a great number of specialized vulnerability prediction models in the literature.
Existing solutions can be divided into three main groups based on their training data: software metrics, text mining, and crash dump stack traces based methods.

The closest to our work is obviously that of Ferenc et al.~\cite{raise}, who created a vulnerability dataset from 12,125 JavaScript functions mapping public vulnerability database entries to code.
They calculated static source code metrics for these functions and trained different machine learning models to predict vulnerable functions.
Our work is heavily related to theirs as we reused their published dataset.
However, instead of just using static source code metrics, we augmented the data with process metrics mined from the version control history of the projects and revisited ML model performances.

In the work by Shin et al.~\cite{5560680}, the researchers performed two empirical case studies on two large, widely used open-source projects: the Mozilla Firefox web browser and the Red Hat Enterprise Linux kernel.
They have used three types of \textbf{software metrics} as prediction features: Code Complexity, Code Churn, and Developer Activity.
Code Complexity belongs to the static source code metrics (included in the metrics we use in this paper as well), Code Churn metrics are types of process metrics, while Developer Activity features describe the relationship between each developer and the project, like the CNCloseness feature, which has high value if a file was modified by developers who are focused on many other files.
They used the tool \textit{Understand\ C++}\footnote{https://www.scitools.com/} for extracting static source code metrics.
They trained five different machine learning models: Logistic Regression, Decision Tree, Random Forest, Naive Bayes, and Bayesian Network.
Their best performing model was Logistic Regression.
This method can predict vulnerabilities in C++ files at file-level, i.e. it can tell whether a \textit{file} contains a vulnerability or not.

Zimmermann et al.~\cite{5477059} trained models on five types of \textbf{software metrics}: Code Complexity, Code Churn, Dependency Measures, Code coverage Measure, and Organizational Measures.
These metrics were calculated at file level and also at the level of binaries.
They trained Logistic Regression on the file-level code metrics and SVM on the binary-level dependency measures.
In their experiments they observed that classical software measures predict vulnerabilities with a high precision but low recall values.
The actual dependencies, however, predict vulnerabilities with a lower precision but substantially higher recall.
We observed similar effect with process metrics, which primarily improved the recall values of the prediction models.
However, they did not cause precision values to decrease.

In the work of Scandariato et al.~\cite{6860243}, the granularity of the created vulnerability prediction model is at the file level, and they analyzed Android programs written in Java.
They used 20 Android applications to build a training dataset.
Their method is based on a \textbf{text mining} technique, which starts with tokenizing each Java file into a vector of terms, and counting the frequency of each term in the file.
Then these values are transformed into a feature vector.
They have initially explored five, well-known learning techniques: Decision Trees, k-Nearest Neighbor, Naıve Bayes, Random Forest, and SVM.
In this initial experiment they discovered that the best results are obtained with Naıve Bayes and Random Forest, so their work focused on these two algorithms only.

Theisen et al. applied~\cite{7202964} a \textbf{stack trace} based method for vulnerability prediction.
They used crash dump stack traces of the Windows 8 operating system.
They parsed the stack traces and their components into a graph representation, then combined multiple stack traces into a single graph.
With this technique, their goal was to identify the software's attack surface, i.e. the sum of all paths where untrusted data can get in and out of the system.
After the identification, they ran their prediction model only on those source files and binaries that belong to the attack surface.
They used the Random Forest machine learning technique to build the prediction model.
The training dataset consisted of features from source code and binary level software metrics data.

The study of Sultana and Williams~\cite{8100852} explores the performance of class-level patterns (i.e., micro patterns) in vulnerability prediction and compares them with traditional class-level software metrics.
They found that micro patterns have higher recall in detecting vulnerable classes than the software metrics.

In contrast to these approaches, we used 42 different static source code metrics and 19 different process metrics as predictor features.
The prediction's granularity is at function level, i.e. our models can predict whether a \textit{function} contains a vulnerability or not, which is much more fine-grained than file-level approaches.
This allows the models to produce more actionable results.
Also, we trained ten different Machine Learning algorithms using our dataset, including deep neural networks and other classical approaches, like Random Forest or k-Nearest Neighbor.

\vspace{-25pt}
\section{\uppercase{Approach}}
\label{sec:approach}
\vspace{-5pt}

\begin{figure*}[t]
  \centering
  \includegraphics[width=0.85\textwidth]{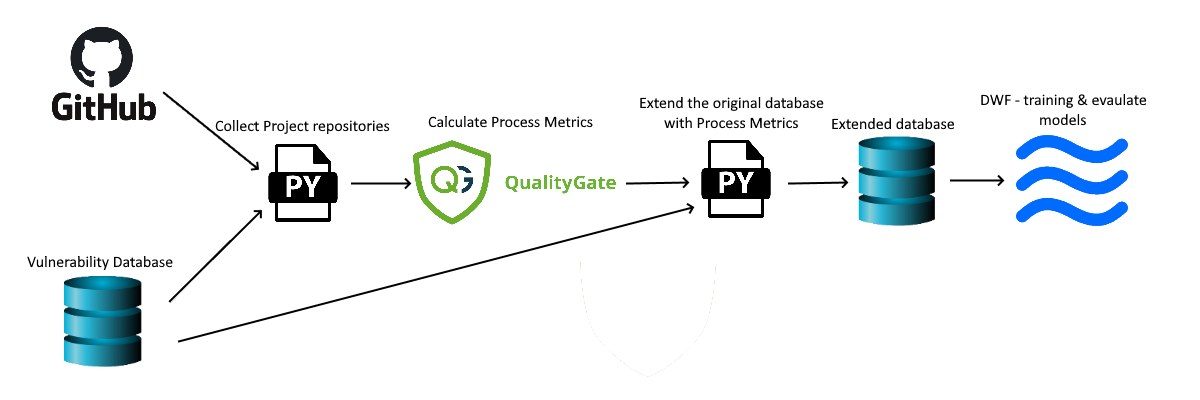}
  \caption{Data extraction and model training approach}
	\vspace{-10pt}
  \label{fig:process}
 \end{figure*}

In this section, we summarize the overall approach for the process metric collection, dataset extension and model training on this extended dataset.
Figure~\ref{fig:process} shows this process at high level.

\vspace{-5pt}
\subsection{The Vulnerability Dataset}
\vspace{-5pt}

As we already pointed out, vulnerabilities can differ from bugs in many cases.
They might not even prevent our program from working properly (rather it allows users to do things they should not be able to).
Therefore, in order to train specific prediction models, we need a dataset built from real-world vulnerabilities in existing code.

To this end, we reused an existing function-level dataset (i.e., functions and their possible vulnerabilities) created by Ferenc et al.~\cite{raise}.
The dataset was generated from functions in JavaScript projects.
To create this vulnerability dataset, the authors collected vulnerability fix commits from GitHub based on two publicly available vulnerability databases, the Node Security Platform (nsp)\footnote{\url{https://github.com/nodesecurity/nsp}} and the Snyk Vulnerability Database.\footnote{\url{https://snyk.io/vuln}}
They mapped the vulnerability fix patches to functions and collected all those that were affected by the patch.
The functions from the code version before the fix became the vulnerable samples, while all other (not affected) functions form the non-vulnerable samples.

After the data collection, they calculated 42 static source code metrics (see Table~\ref{tab:static_metrics} in the appendix) with two different static analysis tools for each analyzed function, like Cyclomatic Complexity, Nesting Level, etc.
This dataset contains function level data from more than 300 projects, and every data row corresponding to a JavaScript function is labeled vulnerable or not vulnerable (labels of $1$ and $0$).
For more details on the dataset collection, please refer to~\cite{raise}.

Using this dataset with the set of static source code metrics as features, Ferenc et al. trained models that achieved an F-measure of 76\% (91\% precision and 66\% recall), which already sounds very promising and provides a strong basis for further research.
In this work, we extended this database with 19 different process metrics (see Table~\ref{tab:process_metrics} in the appendix) and studied how the performance of vulnerability prediction models change.

\subsection{Process Metrics Extraction}\label{sec:pm-desc}

To extend the above mentioned vulnerability dataset with process metrics, we used the algorithm of \textit{Gyimesi et al.}~\cite{peti_proc_met}.
We integrated this algorithm into a software quality monitoring platform, called \textit{QualityGate}\footnote{https://quality-gate.com/}~\cite{bakota2014qualitygate}, which is able to analyze thousands of code versions efficiently, which is a must for calculating process metrics.
The platform loads the \textit{git}\footnote{https://git-scm.com/} version control data of a project as input, performs static analysis on it, and stores in an internal graph format, then calculates process metrics on it.
Therefore, one needs to provide the version control URL for each project, and optionally a specific commit hash, until which they want to calculate the process metrics.
For each program revision the process metrics are updated based on the previous values and the current commit.
So to extend the original dataset with process metrics for each function, we went through the original database, collected the necessary project information, and ran QualityGate to calculate the process metrics.

The output of the algorithm is one csv file per project, which contains the metrics and other information (e.g., line info), which we could use to merge these metrics back into the original dataset as new feature columns.
This way we have extended the dataset with 19 new features, like number of modifications, number of contributors, etc.
More details of the extracted process metrics can be found in~\cite{peti_proc_met}.
To facilitate reproducibility, we made the extended dataset publicly available.\footnote{\url{https://doi.org/10.5281/zenodo.4590021}}

\subsection{Training Configurations}
In order to test the impact of the newly added process metrics on the performance of machine learning models, we trained 10 different algorithms using the Deep-Water Framework\footnote{https://github.com/sed-inf-u-szeged/DeepWaterFramework}~\cite{FERENC2020100551}.
This tool simplifies the configuration management of ML tasks and offers efficient distributed training capabilities and comprehensive visualization features.
In the following subsections, we describe the details of data pre-processing and learning.

\subsubsection{Balancing the Dataset}

The vulnerability dataset is highly imbalanced, as it contains 12,125 functions from which only 1,496 are vulnerable.
Therefore, in addition to the hyperparameters of each model, we included re-sampling parameters in the grid-search we performed to find optimal values.
We tried both over-sampling and under-sampling.
In classification, over-sampling means that we duplicate some of the elements of the less populated class, while under-sampling means that we delete some of the elements of the more populated class.
The result of these two operations is a more balanced dataset, which drastically improves model performances.
If our model was trained on a dataset containing e.g., mostly not vulnerable code, it would make it less likely to flag a particular snippet of code as vulnerable.
The cause of this is that the model will have a ``preconception'' that the largest portion of program code are not vulnerable, since that is what it experienced in the learning phase.

\subsubsection{Used ML Algorithms}\label{sec:ml-algs}
\vspace{-5pt}

We compared the results of all of the available machine learning algorithms that Deep-Water Framework offers by default.
These were the Random Forest Classifier (RFC), Decision Tree Classifier (DT), K-Nearest Neighbor Classifier (KNN), Support Vector Machine Classifier (SVM), Linear Regression (LinReg), Logistic Regression (LogReg), Gaussian Naive Bayes (NB), and Dummy Classifier (ZeroR) from the \textbf{scikit-learn}\footnote{https://scikit-learn.org/} Python library.
The framework also offers two different neural network implementations.
The first one is a default or simple version (SDNN), which works with a fixed learning rate and runs for a fixed number of epochs.
The other one is an improved or complex one (CDNN), which runs validation after every epoch, and if the model gets worse (which is called a miss), then it restores the previous best model and tries again with a reduced (halved) learning rate.
CDNN stops after a given number of misses.
Both of these neural net implementations are based on the Tensorflow\footnote{https://www.tensorflow.org/} Python library.
From now on, we will refer to the algorithms listed above with the abbreviation in parentheses following them.


\subsubsection{Hyperparameter Optimization}
\vspace{-5pt}

Finding the best hyperparameters of a machine learning algorithm is crucial in creating the best possible prediction model.
We used grid-search to find the best parameters, which is a common technique in the field of machine learning.
The name of grid-search comes from how it works:
First, we specify a set of values for each parameter that we want to optimize.
After that, we create the Cartesian product of these sets, train a model using each pair in the resulting set, then we select the best performing one.
If we apply this technique on two parameters, then the Cartesian product of the two sets is a 2D matrix, hence the name grid-search.
We used 10-fold cross validation to evaluate model performances, and applied a train-dev-test split of the dataset with a 80\%-10\%-10\% division.

\vspace{-10pt}
\section{\uppercase{Results}}
\label{sec:results}

In this section, we present our findings about the impact of process metrics on vulnerability prediction.
First, we recap the original performance measures achieved by Ferenc et al. using only static source code metrics as predictors.
Then, we showcase how the addition of process metrics changes the machine learning results.

\vspace{-5pt}
\subsection{Vulnerability Prediction with Static Metrics Only}
\vspace{-5pt}

We used the results published by Ferenc et al. on the original JavaScript vulnerability dataset (containing only static source code metrics) as a baseline for evaluating the effect of the extracted process metrics on the vulnerability prediction.
They achieved the best F-measure value with the KNN algorithm without re-sampling.

\begin{table}[htb!]
\caption{Best results with only static source code metrics achieved by Ferenc et al. (our baseline)}
\label{tab:static_only}
\centering
\begin{tabular}{|c|c|c|c|}
\hline
Classifier & Precision    & Recall       & F-measure  \\
\hline
KNN        & 90.9\%       & 65.9\%       & {\bf 76.4\%} \\
DT         & 73.7\%       & {\bf 69.7\%} & 71.6\%       \\
RFC        & 93.1\%       & 57.8\%       & 71.3\%       \\
SDNN       & 87.3\%       & 60.0\%       & 71.1\%       \\
CDNN       & 91.1\%       & 57.9\%       & 70.8\%       \\
SVM        & {\bf 95.3\%} & 51.4\%       & 66.8\%       \\
LogReg     & 75.3\%       & 21.2\%       & 33.1\%       \\
LinReg     & 84.3\%       & 15.4\%       & 26.1\%       \\
NB         & 22.4\%       & 11.7\%       & 15.3\%       \\
\hline
\end{tabular}
\end{table}

The complete results are shown in Table~\ref{tab:static_only}, the best value in each column is shown in bold, and the rows are ordered by F-measure in a descending order.

\begin{table*}[!htbp]
\caption{Results with the best hyperparameters defined in previous work by Ferenc et al. using only static metrics as predictors}
\label{default_values}
\footnotesize
\centering
\begin{tabular}{|c|c|c|c|c|c|c|c|c|}
\hline
Classifier & TP & TN & FP & FN  & Accuracy & Precision & Recall & F-measure\\
\hline
RFC    & 699 & 7054 & 24  & 261 & {\bf 96.5\%} & 96.7\% & 72.8\% & {\bf 83.1\% (+11.8\%)}\\
DT     & {\bf 723} & 7006 & 72  & {\bf 237} & 96.2\% & 90.9\% & {\bf 75.3\%} & 82.4\% (+10.8\%)\\
CDNN   & 685 & 7027 & 51  & 275 & 95.9\% & 93.1\% & 71.4\% & 80.8\% (+10\%)\\
SDNN   & 665 & 7037 & 41  & 295 & 95.8\% & 94.2\% & 69.3\% & 79.8\% (+8.7\%)\\
KNN    & 613 & 7059 & 19  & 347 & 95.5\% & {\bf 97.0\%} & 63.9\% & 77.0\% (+0.6\%)\\
SVM    & 548 & {\bf 7060} & {\bf 18}  & 412 & 94.7\% & 96.8\% & 57.1\% & 71.8\% (+5\%)\\
LogReg & 332 & 7007 & 71  & 628 & 91.3\% & 82.4\% & 34.6\% & 48.7\% (+15.6\%)\\
LinReg & 274 & 7051 & 27  & 686 & 91.1\% & 91.0\% & 28.5\% & 43.5\% (+17.4\%)\\
NB     & 115 & 6779 & 299 & 845 & 85.8\% & 27.8\% & 12.0\% & 16.7\% (+1.4\%)\\
\hline
\end{tabular}
\vspace{-11pt}
\end{table*}

\subsection{Vulnerability Prediction with Both Static and Process Metrics}
\vspace{-6pt}

We present how the addition of process metrics changed the results of the prediction models.
We start with outlining the performance of the models using the parameters determined in the previous work of Ferenc et al., then we detail how re-sampling and hyperparameter optimization affect the results on the extended dataset.
Finally, we compare the best models achieved by the addition of process metrics, and the best models relying only on static code metrics.

\vspace{-5pt}
\subsubsection{Best Results with Hyperparameters from Previous Work}
\vspace{-8pt}

We trained 10 different models (with ML algorithms described in Section~\ref{sec:ml-algs}) on the dataset extended with process metrics (see Section~\ref{sec:pm-desc}).
We included the ZeroR classifier as well, which classifies everything based on the most common label in the dataset (i.e., we used it as a baseline algorithm).
As the majority of samples fall into the non-vulnerable category (i.e., 0/negative class label), ZeroR has no precision and a 0 recall measure, but it set up a baseline for accuracy, 86.93\%.
This means that if another model performs worse in terms of accuracy than ZeroR, then it is no better than the trivial guessing.

We continued with the evaluation of the remaining 9 models.
First, we used the best hyperparameters determined in the previous work by Ferenc et al. for each algorithm (using only static source code metrics) to get an overall picture of the performance of these models on the extended, new dataset.
As shown in Table~\ref{default_values}, the best overall performance, i.e. F-measure, was achieved by RFC in this setup.
This is 6.7\% higher than the best model (KNN) based only on the static source code metrics.
The change in F-measure of each model compared to its version based on static source code metrics only is shown in parentheses.

The overall accuracy of the RFC model was 96.5\%, which is almost 10\% higher than that of the trivial ZeroR algorithm.
The effect on process metrics could be observed on both precision and recall, which improved at almost the same rate.
The worst result in this setup was achieved by NB, which underperformed even ZeroR (but still outperformed NB using only static source code metrics).
Based on these first results, we can already answer RQ1 as follows.

\vspace{5pt}
\noindent\fbox{\parbox{0.46\textwidth}{
\textbf{RQ1}: Process metrics can improve existing JavaScript vulnerability prediction models.
With this initial experiment, using the best hyperparameters from the previous work, we already improved the best F-measure value by 6.7\%, the best recall by 5.6\%, and the best precision by 1.7\%.
}}
\vspace{2pt}

Another important factor is the number of false positives (FP column in Table~\ref{default_values}), where the best value was achieved by SVM, KNN being the second best.
Although KNN and SVM did not perform as well as RFC in terms of F-measure, they produced the lowest number of false hits, while having the two highest precision values.
That is, there may be vulnerabilities that the model does not recognize, but in return, the number of false alarms they produce is significantly reduced.
In terms of usability, this might be an important aspect, as false alarms can easily build distrust in the users, preventing the adoption of such models in practice.

\begin{table*}[!htbp]
\caption{Best F-measure values per classifier using the optimal hyperparameters}
\label{best_fmeasure}
\footnotesize
\centering
\begin{tabular}{|c|c|c|c|c|c|c|c|c|}
\hline
Classifier & TP & TN & FP & FN & Accuracy & Precision & Recall & F-measure\\
\hline
RFC    & {\bf 730} & {\bf 7046} & {\bf 32} & {\bf 230} & {\bf 96.7\%} & {\bf 95.8\%} & {\bf 76.0\%} & {\bf 84.8\%} {\bf (+13.5\%)} \\
DT     & 723 & 7006 & 72  & 237 & 96.2\% & 90.9\% & 75.3\% & 82.4\% (+10.8\%)\\
KNN    & 684 & 7041 & 37  & 276 & 96.1\% & 94.9\% & 71.3\% & 81.4\% (+5\%) \\
SDNN   & 687 & 7019 & 59  & 273 & 95.9\% & 92.1\% & 71.6\% & 80.5\% (+9.4\%) \\
CDNN   & 678 & 7025 & 53  & 282 & 95.8\% & 92.8\% & 70.6\% & 80.2\% (+9.4\%)  \\
SVM    & 692 & 6966 & 112 & 268 & 95.3\% & 86.1\% & 72.1\% & 78.5\% (+11.7\%) \\
LogReg & 496 & 6906 & 172 & 464 & 92.1\% & 74.3\% & 51.7\% & 60.9\% (+27.8\%)   \\
LinReg & 570 & 6592 & 486 & 390 & 89.1\% & 54.0\% & 59.4\% & 56.6\% (+24.5\%) \\
NB     & 115 & 6779 & 299 & 845 & 85.8\% & 27.8\% & 12.0\% & 16.7\% (+1.4\%) \\
\hline
\end{tabular}
\vspace{-10pt}
\end{table*}

\subsubsection{The Effects of Re-Sampling}
\vspace{-5pt}

To further improve the efficiency of the models on the extended dataset, we used four different re-sampling ratios, both for under and over-sampling: 25\%, 50\%, 75\%, 100\%.
So we used under and over-sampling strategies to reach the ratio of positive and negative samples to be at these levels.
We observed that by performing either under or over-sampling to achieve the same ratios, we get similar model performances.
With higher re-sampling rates a higher recall value can be achieved, which means the models could recognize a higher percentage of vulnerable functions.
Unfortunately, the higher percentage of recognized vulnerable functions is accompanied by a higher number of false positive elements, i.e. the model is more likely to label a function to be vulnerable that does not actually contain any vulnerabilities.
For all the other performance metrics, the best results were obtained without re-sampling, therefore, in each table we present the results achieved on the dataset without re-sampling.

\subsubsection{Finding the Best Hyperparameters}
\vspace{-5pt}

To find out if further improvements could be achieved by hyperparameter optimization based on the extended new dataset, we started a grid-search for each classifier one by one to find the best possible parameterizations.
Each row in Table~\ref{best_fmeasure} shows the best result in terms of F-measure achieved by the grid-search on a classifier's hyperparameters.
The table is organized by F-measure in a descending order for easier overview.

As we can see, hyperparameter optimization on the extended dataset further increased the performance measures.
The magnitude of this increase is varying, but can be as high as +12.2\% for LogReg.
The best results are shown in bold; the F-measure improvement compared to the static source metrics results are also shown in the last column, between parentheses.
Again, RFC performed by far the best, but DT, KNN, and neural network-based methods also worked well.
What is remarkable is that RFC and KNN achieved a fairly high precision value despite the results being sorted by F measure.
This is important because even though we want to maximize the precision value, the recall -- thus the overall performance -- is also very important, as a poorly performing model in this respect can give a false sense of security to the users.

Each row in Table~\ref{best_precision} shows the best result in terms of precision achieved by the grid-search on a classifier's hyperparameters.
The table is organized by precision in a descending order for easier overview.
The best results are shown in bold here as well.
Again, the aforementioned decision tree-based models (DT, RFC) performed among the best, but KNN achieved the best precision value.

As we mentioned earlier, on the one hand, maximizing the precision value is important in terms of increasing user experience.
On the other hand, F-measure is still important, since too many false negatives can lead to a false sense of security, which can also result in bad user experience.
We can see that DT achieved a higher precision value than RFC, but DT performed significantly worse than RFC in terms of F-measure (i.e., has much more FN hits, thus lower recall).
Furthermore, selecting hyperparameters that optimize model precision significantly reduces their recall, thus the overall F-measure.
So we could confirm that there is a practical trade-off between precision and recall, which one needs to consider when selecting the hyperparameters of models for practical applications.

\begin{table*}[!htbp]
\caption{Best precision values per classifier using the optimal hyperparameters}
\label{best_precision}
\footnotesize
\centering
\begin{tabular}{|c|c|c|c|c|c|c|c|c|}
\hline
classifier & TP & TN & FP & FN & accuracy & precision & recall & F-measure\\
\hline
KNN   & 565 & {\bf 7071} & {\bf 7} & 395 & 95.0\% & {\bf 98.8\%} & 58.9\% & 73.8\% \\
DT     & 429 & 7068 & 10  & 531 & 93.3\% & 97.7\% & 44.7\% & 61.3\% \\
RFC    & {\bf 599} & 7063 & 15 & {\bf 361} & {\bf 95.3\%} & 97.6\% & {\bf 62.4\%} & {\bf 76.1\%} \\
SVM    & 548 & 7060 & 18  & 412 & 94.7\% & 96.8\% & 57.1\% & 71.8\% \\
CDNN   & 551 & 7048 & 30  & 409 & 94.5\% & 94.8\% & 57.4\% & 71.5\% \\
SDNN   & 572 & 7045 & 33  & 388 & 94.8\% & 94.6\% & 59.6\% & 73.1\% \\
LogReg & 221 & 7058 & 20  & 739 & 90.6\% & 91.7\% & 23.0\% & 36.8\% \\
LinReg & 274 & 7051 & 27  & 686 & 91.1\% & 91.0\% & 28.5\% & 43.5\% \\
NB     & 115 & 6779 & 299 & 845 & 85.8\% & 27.8\% & 12.0\% & 16.7\% \\
\hline
\end{tabular}
\vspace{-10pt}
\end{table*}

\subsubsection{Comparison of the Best Models}
\vspace{-5pt}

\begin{figure}[!htbp]
  \centering
  \includegraphics[width=\columnwidth]{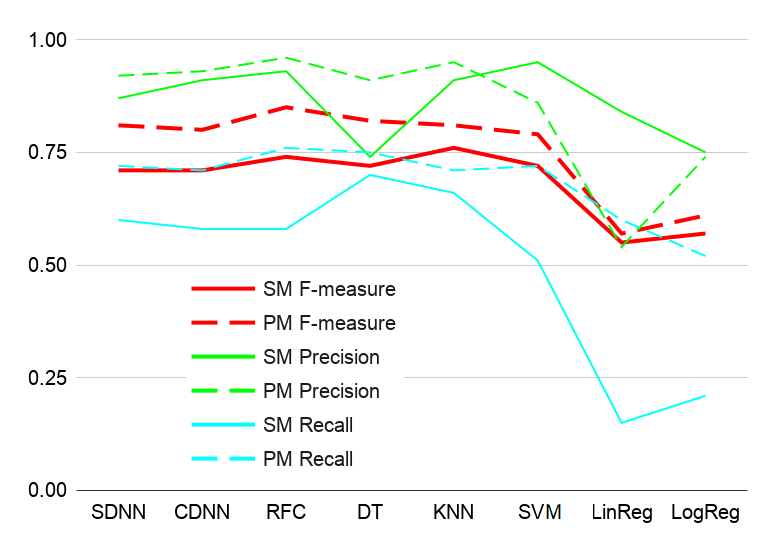}
  \caption{Overview of the best precision, recall, and F-measure values for models using static metrics (SM) only, and models using static and process metrics (PM) combined}
  \label{fig:results}
\end{figure}

To aid easier comparison of the data presented in the tables above, we created an overview plot of the best results of each algorithm (see Figure~\ref{fig:results}).
Here the best results per algorithm can be seen, according to three different metrics, F-measure (or F1-score), precision, and recall.
Recall measures what percentage of the vulnerable functions was found by our model, precision measures what percentage of the predicted vulnerable functions was truly vulnerable, while F-measure is the harmonic mean of these two.

Using this plot, one can easily see how these values changed on the extended dataset.
F-measure values are marked with red color.
The full red line shows the best results with only static source metrics as predictors, and the dashed red line shows the results obtained with with the process metrics being added to the set of features.
We applied the same convention for the other two metrics, precision and recall.

As can be seen, except in the case of SVM, LinReg and LogReg classifiers, we achieved significant improvement in all three metrics.
The recall value was improved the most by process metrics, which means that using these metrics, we can discover a lot more vulnerabilities.
In case of the above mentioned three algorithms, the price of recall improvement was a drop in precision, which also resulted in a less significant overall improvement in terms of F-measure.
Even though intuitively it seems that adding process metrics as features improve ML models in predicting vulnerable JavaScript functions, we need to perform a significance test to be sure that the improvements are not by coincidence.

\vspace{-4pt}
\subsubsection{Checking the Statistical Significance of Performance Improvement}
\vspace{-4pt}

To check if the improvements are statistically significant, we applied the McNemar's test~\cite{lachenbruch2014mcnemar}.
This test is suitable for comparing the prediction power of two classifiers.
The default assumption, or null hypothesis, of the test is that the two classifiers disagree to the same amount.
If the null hypothesis is rejected, it suggests that there is evidence that the cases disagree in different ways, i.e., the disagreements are skewed.
In our case, we took one of two classifiers (for every algorithm pair) trained on the original database, which only contains static metrics, while the other one trained on the new, extended database, which contains process metrics too.
The McNemar’s test operates upon a 2x2 contingency table, which we can produce by evaluating both models on the same dataset (in our case, these two datasets were different in terms of the number of features, but both models were predicting the same functions, so we satisfied this requirement).
The contingency table is produced as follows.
At the 0,0 position in the table, we count the number of occurrences where both classifiers predicted the correct label, and at the 1,1 index we count the number of occurrences where both classifiers predicted the incorrect label.
At 0,1 we count the number of occurrences where the first classifier predicted the correct label, but the second classifier predicted the wrong label, and vice versa at 1,0.
The McNemar's test calculates a so called ``statistic'' value as follows:
$$
statistic = \frac{(ct[0][1] - ct[1][0])^{2}}{(ct[0][1] + ct[1][0])},
$$
where $ct[x][y]$ is the value of contingency table at position (x, y).
Using this statistic value, the test calculates a \textit{p} value, which can be interpreted as follows.
\textbf{p \textgreater  $\alpha$}: accept H0, e.g. there is no significant improvement.
\textbf{p \textless= $\alpha$}: reject H0, e.g. the improvement in prediction performances is significant.
We used the Statsmodels~\footnote{https://www.statsmodels.org/stable/index.html} Python library to implements the test.
The results are summarized in Table~\ref{tab:mcnemar}.
As it is shown in the table in case of all the used algorithms, the improvement was statistically significant (we used 0.05 $\alpha$ value to reject the null hypothesis).

\begin{table}[!htbp]
\caption{McNemar's test}
\label{tab:mcnemar}
\centering
\begin{tabular}{|c|c|c|c|}
\hline
\textbf{Model} & \textbf{Statistic} & \textbf{p-value} & \textbf{H0} \\
\hline
SDNNC  & 27.000  & 0.000 & rejected \\
CDNNC  & 55.000  & 0.000 & rejected \\
DT     & 44.000  & 0.000 & rejected \\
RFC    & 18.000  & 0.000 & rejected \\
LinReg & 10.000  & 0.000 & rejected \\
LogReg & 28.000  & 0.000 & rejected \\
LogReg & 28.000  & 0.000 & rejected \\
NB     & 131.000 & 0.048 & rejected \\
SVM    & 12.000  & 0.000 & rejected \\
KNN    & 12.000  & 0.000 & rejected \\
\hline
\end{tabular}
\vspace{-5pt}
\end{table}

After the results of this McNemar's test, we are ready to formulate our answer to RQ2.

\vspace{5pt}
\noindent\fbox{\parbox{0.46\textwidth}{
\textbf{RQ2}: We can achieve significant improvement in predicting vulnerable JavaScript functions using process metrics as additional features. The overall performance increase with process metrics was 8.4\% in terms of F-measure (in the baseline work, Ferenc et al. achieved an F-measure of 0.76 with KNN, now on the new dataset we achieved an F-measure of 0.85 with RFC), 6.3\% in terms of recall and a 3.5\% in terms of precision. All the performance measure increases are statistically significant.
}}
\vspace{-15pt}
\section{Threats to Validity}\label{sec:threats}
\vspace{-10pt}


The complete dataset we trained our models contains only 12,125 samples.
Some models, typically deep neural networks require a much higher number of samples to produce reliable results.
However, we applied classical machine learning models, too to build vulnerability prediction models, which were performing even better than the deep neural networks.
Therefore, we are certain about our findings, nonetheless, further extension of the dataset would be favorable.

The training dataset is highly imbalanced.
Less then 10\% of the samples were actually vulnerable.
To mitigate this risk, we applied re-sampling strategies before model training.
However, we found that most of the models are affected by re-sampling in a controversial way, thus we presented the results without applying any re-sampling on the data.

There are several ML and evaluation measures, and we used precision, recall and F1-measure for comparison.
However, for highly imbalanced datasets, AUC, probability of false alarms, recall , G-mean, or MCC might work better.
Nonetheless, we believe the general conclusions would remain valid based on these measures as well.
\vspace{-15pt}
\section{\uppercase{Conclusion}}\label{sec:conclusion}
\vspace{-10pt}

In this paper, we studied the effect of process metrics on the prediction performance of ML models in the context of JavaScript functions. We based our research on a previously published vulnerability dataset by Ferenc et al.
The dataset contained 42 static source code metrics of various JavaScript functions and a label column indicating whether a function contains a vulnerability or not.
We extended this dataset with 19 process metrics, which was calculated by mining the version control history of each project.

Using the data set, we trained different machine learning models to predict vulnerable functions.
We compared the performance of these models with the ones obtained by using only static source code metrics as predictors.
The best performing algorithm for predicting vulnerable JavaScript functions relying on the combination of process metrics and static source code metrics was Random Forest in terms of F-measure with a value of 0.85 (0.96 precision and 0.76 recall), which is a clear improvement compared to the best model (k-NN) using only static source code metrics that achieved an F-measure of 0.76 (0.91 precision and 0.66 recall).

The best precision (0.99) was achieved by k-NN (with the static metrics it was 0.95), while the best recall (0.90) was achieved by SVM (with the static metrics it was 0.80).
In overall, according to our experiences, Random Forest and k-NN are mostly equally well-suited for the task, while the regressions as well as the Naive Bayes algorithm performed worse.

The results showed that using process metrics can significantly improve the predictive power of JavaScript vulnerability prediction models in terms of all the IR measures (i.e., precision, recall, F-measure).
With a McNemar's test, we also showed that the increase in performance measures are statistically significant.
In the future, we plan to extend the feature space with some new type of metrics and make such prediction models part of the developer's everyday lives.

\vspace{-15pt}
\section*{\uppercase{Acknowledgment}}
\vspace{-5pt}
The presented work was carried out within the SETIT Project (2018-1.2.1-NKP-2018-00004)\footnote{Project no. 2018-1.2.1-NKP-2018-00004 has been implemented with the support provided from the National Research, Development and Innovation Fund of Hungary, financed under the 2018-1.2.1-NKP funding scheme.} and supported by the Ministry of Innovation and Technology NRDI Office within the framework of the Artificial Intelligence National Laboratory Program (MILAB). The research was partly supported by the EU-funded project AssureMOSS (Grant no. 952647).

Furthermore, Péter Hegedűs was supported by the Bolyai János Scholarship of the Hungarian Academy of Sciences and the ÚNKP-20-5-SZTE-650 New National Excellence Program of the Ministry for Innovation and Technology.

\bibliographystyle{apalike}
{\small
\vspace{-15pt}
\bibliography{references,bibl}}

\section*{\uppercase{Appendix}}
\label{sec:appendix}

\begin{table}[h]
\vspace{-5pt}
\caption{Process Metrics}
\label{tab:process_metrics}
\centering
\resizebox{\columnwidth}{!}{%
\begin{tabular}{|c|c|}
\hline
\textbf{Metric} & \textbf{Description} \\
\hline
AVGNOAL         & Average Number Of Added Lines                \\
AVGNODL         & Average Number Of Deleted Lines              \\
AVGNOEMT        & Average Number Of Elements Modified Together \\
AVGNOML         & Average Number of Modified Lines             \\
AVGTBC          & Average Time Between Changes                 \\
CChurn          & Sum of lines added minus lines deleted       \\
MNOAL           & Maximum Number of Added Lines                \\
MNODL           & Maximum Number of Deleted Lines              \\
MNOEMT          & Maximum Number of Elements Modified Together \\
MNOML           & Maximum Number of Modified Lines             \\
NOADD           & Number of Additions                          \\
NOCC            & Number of Contributor Changes                \\
NOCHG           & Number of Changes                            \\
NOContr         & Number of Contributors                       \\
NODEL           & Number of Deletions                          \\
NOMOD           & Number of Modifications                      \\
SOADD           & Sum of Added Lines                           \\
SODEL           & Sum of Deleted Lines                         \\
SOMOD           & Sum of Modified Lines                        \\
\hline
\end{tabular}%
}
\vspace{-15pt}
\end{table}

\begin{table}[h]
\caption{Static Metrics}
\label{tab:static_metrics}
\centering
\resizebox{\columnwidth}{!}{%
\begin{tabular}{|c|c|}
\hline
\textbf{Metric} & \textbf{Description} \\
\hline
CC              & Clone Coverage                     \\
CCL             & Clone Classes                      \\
CCO             & Clone Complexity                   \\
CI              & Clone Instances                    \\
CLC             & Clone Line Coverage                \\
LDC             & Lines of Duplicated Code           \\
McCC, CYCL      & Cyclomatic Complexity              \\
NII             & Number of Incoming Invocations     \\
NL              & Nesting Level                      \\
NLE             & Nesting Level without else-if      \\
NOI             & Number of Outgoing Invocations     \\
CD, TCD         & (Total) Comment Density            \\
CLOC, TCLOC     & (Total) Comment Lines of Code      \\
DLOC            & Documentation Lines of Code        \\
LLOC, TLLOC     & (Total) Logical Lines of Code      \\
LOC, TLOC       & (Total) Lines of Code              \\
NOS, TNOS       & (Total) Number of Statements       \\
NUMPAR, PARAMS  & Number of Parameters               \\
HOR\_D          & Nr. of Distinct Halstead Operators \\
HOR\_T          & Nr. of Total Halstead Operators    \\
HON\_D          & Nr. of Distinct Halstead Operands  \\
HON\_T          & Nr. of Total Halstead Operands     \\
HLEN            & Halstead Length                    \\
HVOC            & Halstead Vocabulary Size           \\
HDIFF           & Halstead Difficulty                \\
HVOL            & Halstead Volume                    \\
HEFF            & Halstead Effort                    \\
HBUGS           & Halstead Bugs                      \\
HTIME           & Halstead Time                      \\
CYCL\_DENS      & Cyclomatic Density                 \\
WarningInfo     & ESLint Info priority Warnings      \\
WarningMinor    & ESLint Minor priority Warnings     \\
WarningMajor    & ESLint Major priority Warnings     \\
WarningCritical & ESLint Critical priority Warnings  \\
WarningBlocker  & ESLint Blocker priority Warnings   \\
\hline
\end{tabular}%
}
\end{table}

\end{document}